\documentclass[12pt,reqno,a4paper]{amsart}
\usepackage [latin1]{inputenc}
\usepackage{amscd,amsfonts,amssymb,amsmath,amsthm}
\usepackage[arrow,matrix]{xy}
\usepackage{color}
\usepackage{mathrsfs,euscript,hyperref}

\newtheorem{thm}{Theorem}[section]
\theoremstyle{remark}

\theoremstyle{definition}
\newtheorem{definition}{Definition}[section]

\newcommand{\xbar}[1]{%
   \kern0.15ex\hbox{%
     \vbox{%
       \hrule height 0.4pt 
       \kern0.3ex
       \hbox{%
         \kern-0.15em
         \ensuremath{#1}%
         \kern-0.1em
       }%
     }%
   \kern0.25ex}%
}

\newcommand{\xbarscr}[1]{%
   \kern0.2ex\hbox{%
     \vbox{%
       \hrule height 0.3pt 
       \kern0.2ex
       \hbox{%
         \kern-0.1em
         \ensuremath{{}_{#1}}%
         \kern-0.15em
       }%
     }%
   \kern0.25ex}%
}

\newcommand{\xbarscrscr}[1]{%
   \kern0.2ex\hbox{%
     \vbox{%
       \hrule height 0.3pt 
       \kern0.2ex
       \hbox{%
         \kern-0.1em
         \ensuremath{{}_{{}_{#1}}}%
         \kern-0.15em
       }%
     }%
   \kern0.25ex}%
}

\newcommand{\checkbxi}{\lefteqn{\boldsymbol{\xi}}\kern.13pc\check{\phantom{\xi}}\kern-.09pc}

\newtheorem{pro}[section]{Proposition}

\numberwithin{equation}{section}

\begin{document}

\title[GRADIENT GIBBS MEASURES FOR THE SOS MODEL]{
GRADIENT GIBBS MEASURES FOR THE SOS MODEL WITH
INTEGER SPIN VALUES ON A CAYLEY TREE}
\author{G. I. Botirov,  \,  F. H. Haydarov}

\address{G. Botirov, Institute of Mathematics, Academy of Science of Uzbekistan,
Tashkent, Uzbekistan,}
\email{botirovg@yandex.ru}

 \address{F.Haydarov, National University of Uzbekistan,
Tashkent, Uzbekistan}
\email{haydarov\_imc@mail.ru}
\vspace{10pt}

\begin{abstract}
In the present paper we continue the investigation from [1] and consider the SOS (solid-on-solid) model on the Cayley tree of order $k \geq 2$. In the ferromagnetic SOS case on the Cayley tree, we find three solutions to a class of period-4 height-periodic boundary law equations and these boundary laws  define up to three periodic gradient Gibbs measures.
\end{abstract}
\maketitle


\section{Introduction}

A solid-on-solid (SOS) model can be considered as a generalization of the Ising model, which corresponds to
$E=\{-1,1\}$, or a less symmetric variant of the Potts model with non-compact state space.
SOS-models on the cubic lattice were analyzed in \cite{Maz},\cite{BW} where an analogue of the so-called Dinaburg-Mazel-Sinai theory was developed. Besides interesting phase transitions in these models, the attention to them is motivated by applications, in particular in the theory of communication networks; see, e.g., \cite{BK1}, \cite{G}, \cite{G'}, \cite{Ro}.

In \cite{HK} it is shown that on the Cayley tree there are several tree automorphism invariant gradient Gibbs measures and the existence of $q$ different gradient Gibbs measures for $q$-component models on the Cayley tree of order $k\geq 2$. To the best of our knowledge, the first paper devoted to the SOS model on the Cayley tree is \cite{Ro12}. In \cite{Ro12} the case of arbitrary $m\geq 1$ is treated and a vector-valued functional equation for possible boundary laws of the model is obtained. Recall that each solution to this functional equation determines a splitting Gibbs measure (SGM), in other words a tree-indexed Markov chain which is also a Gibbs measure. Such measures can be obtained by propagating spin values along the edges of the tree, from any site singled out to be the root to the outside, with a transition matrix depending on initial Hamiltonian and the boundary law solution. In particular the homogeneous (site-independent) boundary laws then define translation-invariant (TI) SGMs. For a recent investigation of the influence of weakly non-local perturbations in the interaction
to the structure of Gibbs measures, see \cite{be} in the context of the Ising model.

The present paper is organized as follows.  In Section 2 we present the preliminaries of the model. In the third section we construct gradient Gibbs measures for period 4 height-periodic boundary laws on the Cayley tree of order $k\geq 2$. Note that the results in \cite{HK} are proved only on the Cayley tree of order two.

\section{Preliminaries}

{\it Cayley tree.} The Cayley tree $\Gamma^k$ of order $ k\geq 1 $
is an infinite tree, i.e., a graph without cycles, such that
exactly $k+1$ edges originate from each vertex. Let $\Gamma^k=(V,
L)$ where $V$ is the set of vertices and  $L$ the set of edges.
Two vertices $x$ and $y$ are called {\it nearest neighbors} if
there exists an edge $l \in L$ connecting them. We will use the
notation $l=\langle x,y\rangle$. A collection of nearest neighbor
pairs $\langle x,x_1\rangle, \langle x_1,x_2\rangle,...,\langle
x_{d-1},y\rangle$ is called a {\it path} from $x$ to $y$. The
distance $d(x,y)$ on the Cayley tree is the number of edges of the
shortest path from $x$ to $y$.

For a fixed $x^0\in V$, called the root, we set
\begin{equation*}
W_n=\{x\in V\,| \, d(x,x^0)=n\}, \qquad V_n=\bigcup_{m=0}^n W_m
\end{equation*}
and denote
$$
S(x)=\{y\in W_{n+1} :  d(x,y)=1 \}, \ \ x\in W_n, $$ the set  of
{\it direct successors} of $x$.

{\it SOS model.} We consider a model where the spin takes values in
the set of all integer numbers $\emph{Z}:=\{\dots, -1,0,1,\dots
\}$, and is assigned to the vertices of the tree. A configuration
$\sigma$ on $V$ is then defined as a function $x\in V\mapsto\sigma
(x)\in\emph{Z}$; the set of all configurations is $\Omega:=\emph{Z}^V$.

The (formal) Hamiltonian of the SOS model is :
\begin{equation}\label{nu1}
 H(\sigma)=-J\sum_{\langle x,y\rangle\in L}
|\sigma(x)-\sigma(y)|,
\end{equation}
where $J \in \emph{R}$ is a constant and  $\langle
x,y\rangle$ stands for nearest neighbor vertices.

Note that the Hamiltonian is invariant under the spin-translation/height-shift $t:\left(t \omega \right)_i=\omega_i+t.$
This suggests reducing the complexity of the configuration space by considering {it gradient configurations } instead of height configuartions as will be explained in the following:

{\it Gradient configuration.}  We may induce an orientation on $\Gamma ^k$ relative to an arbitrary site $\rho$ (which we may call the root) by calling an edge $\langle x,y\rangle$ oriented iff it points away from the $\rho$. More precisely, the set of oriented edges is defined by $$  \vec{L}:=\vec{L}_{\rho}:=\{\langle x,y \rangle \in L : d(\rho,y)=d(\rho , x)+1\}.$$
Note that the oriented graph $(V, \vec{L})$ also possesses all tree-properties, namely connectedness and absence of loops.

For any height configuration $\omega =(\omega (x))_{x \in V} \in Z^V$ and $b=\langle x, y \rangle \in \vec{L}$ the height difference along the edge $b$ is given by $ \nabla \omega_b=\omega_y-\omega_x $ and we also call $ \nabla \omega$ the gradient field of $\omega$. The gradient spin variables are now defined by $\eta_{\langle x,y \rangle}=\omega_y-\omega_x$ for each $\langle x,y \rangle \in \vec{L}$. Let us denote the space of {\it gradient configuration} by $\Omega^{\nabla}=Z^{\vec{L}}$. Note that in contrast to the notation used in \cite{S} for the lattice $Z^d$, the gradient configurations defined above are indexed by the oriented edges of the tree and not by its vertices. Equip the integers $Z$ with the power ste as measurable structure. Having done this, the measurable structure on the space $\Omega ^{\nabla}$ is given by the product $\sigma$-algebra $\mathcal{F}^{\nabla}:=\sigma(\{\nabla _ b | b \in \vec{L}\})$. Clearly $\nabla: (\Omega, \mathcal{F}) \rightarrow (\Omega^{\nabla}, \mathcal{F}^{\nabla}) $ then becomes a measurable map.

\section{Gradient Gibbs measures and tree-automorphism invariant solutions}
 \subsection{Gibbs and Gradient Gibbs measures}

 Recall that the set of height configurations $\Omega:=Z^V$ was endowed with the product $\sigma$-algebra $\otimes_{i \in V}2^Z$, where $2^Z$ denotes the power set of $Z$. Then, for any $\Lambda \subset V$, consider the coordinate projection map $\sigma_\Lambda:Z^V \rightarrow Z^{\Lambda}$ and the $\sigma$-algebra $\mathcal{F}_{\Lambda}:=\sigma(\sigma_{\lambda})$ of cylinder sets on $Z^V$ generated by the map $\sigma_{\Lambda}$.

 We define Gibbs measures on the space of height-configurations for the model (\ref{nu1}) on a Cayley tree. Let $\nu =\{\nu(i)>0, \ i \in Z\}$ be $\sigma$-finite positive fixed a-priori measure, which in the following we will always assume to be the counting measure.

 Gibbs measures are built within the DLR framwork by describing conditional probabilities w.r.t. the outside of finite sets, where a boundary condition is frozen. One introduces a so-called Gibbsian specification $\gamma$ so that any Gibbs measure $\mu \in \mathcal{G}(\gamma)$ specified by $\gamma$ verifies
 \begin{equation}
 \mu(A|\mathcal{F}_{\Lambda^c})=\gamma_{\Lambda}(A|\cdot) \ \mu -a.s
 \end{equation}
 for all $\Lambda \in \mathcal{S}$ and $A \in \mathcal{F}$. The Gibbsian specification associated to a potential $\Phi$ is given at any inverse temperature $\beta>0$, for any boundary condition $\omega \in \Omega$ as
 \begin{equation}\label{33}
 \gamma_{\Lambda}(A|\omega)=\frac{1}{Z_{\Lambda}^{\beta, \Phi}}\int e^{-\beta H_{\Lambda}^{\Phi}(\sigma_{\Lambda}\omega_{\Lambda^c})}\mathbf{1}_A (\sigma_{\Lambda}\omega_{\Lambda^c}) \nu^{\otimes \Lambda}(d \sigma_{\Lambda}), \end{equation}
 where the partition function $Z_{\Lambda}^{\beta, \Phi}$-that has to be non-null and convergent in this countable infinite state-space context (this means that $\Phi$ is $\nu$-admissible in the terminology of \cite{Ge})-is the standard normalization whose logarithm is often related to pressure or free energy.

In our SOS-model on the Cayley tree, $\Phi$ is the unbounded nearest-neighbour potential
with $\Phi_{\{x,y\}}(\omega_x-\omega_y)=|\omega_x-\omega_y|$ and $\Phi_{x}\equiv 0$, so $\gamma$
 is a \emph{Markov specification} in the sense
that
\begin{equation}\label{3.6} \gamma_{\Lambda}(\omega_{\Lambda}=\zeta | \cdot)\ \textrm{is}\ \mathcal{F}_{\partial\Lambda}- \textrm{measurable for all} \ \Lambda\subset V \ \ {and}\ \zeta\in \emph{Z}^{\Lambda}. \end{equation}

In order to build up gradient specifications from the Gibbsian specifications defined in \cite{HK}, we need to consider the following: Due to the absence of loops in trees, for any
finite subgraph $\Lambda / \emph{Z}$, the complement $\Lambda^{c}$ is not connected, but consists of at least two
connected components where each of these contains at least one element of $\partial\Lambda$. This
means that the gradient field outside $\Lambda$ does not contain any information on the relative
height of the boundary $\partial\Lambda$ (which is to be understood as an element of $\emph{Z}^{\partial\Lambda}\setminus \emph{Z}$). More
precisely, let $cc(\Lambda^c)$ denote the number of connected components in $\Lambda^c$ and note that $2\leq cc(\Lambda^c)\leq |\partial\Lambda|$.

Applying the general definition of Gradient Gibbs measure (see \cite{HK}) we have

\begin{equation}\label{37} \emph{Z}^{\Lambda^c}/ \emph{Z}=\emph{Z}^{\{b\in\vec{L}\}| b\subset \Lambda^c}\times (\emph{Z}^{cc(\Lambda^c)}/ \emph{Z}\subset \emph{Z}^{\{b\in\vec{L}\}| b\subset \Lambda^c}\times (\emph{Z}^{\partial\Lambda}/ \emph{Z})
\end{equation}
where "=" is in the sense of isomorphy between measurable spaces. For any $\eta\in \Omega^{\nabla}=\emph{Z}^{V}/ \emph{Z}$, let $[\eta]_{\partial\Lambda}/ \emph{Z}$ denote the image of $\eta$ under the coordinate projection
$\emph{Z}^{V}/ \emph{Z}\rightarrow \emph{Z}^{\partial \Lambda}/ \emph{Z}$ with the latter set endowed with the final $\sigma$-algebra generated by the coset projection. Set
\begin{equation}\label{3.8} \mathcal{F}_{\Lambda}^{\nabla}:=\sigma((\eta_b)_{b\subset\Lambda^c})
\subset \mathcal{T}_{\Lambda}^{\nabla}:=\sigma((\eta_b)_{b\subset\Lambda^c}, [\eta]_{\partial\Lambda}).\end{equation}

Then $\mathcal{T}_{\Lambda}^{\nabla}$ contains all information on the gradient spin variables outside $\Lambda$ and also
information on the relative height of the boundary $\partial\Lambda$. By (\ref{37}) we have that for any
event $A\in \mathcal{F}^{\nabla}$ the $\mathcal{F}_{\Lambda^{c}}$-measurable function
$\gamma_{\Lambda}(A | \cdot)$ is also measurable with respect to
$\mathcal{T}_{\Lambda}^{\nabla}$, but in general not with respect to $\mathcal{F}_{\Lambda}^{\nabla}$. These observations lead to the following:
\begin{definition}\label{3.1} The gradient Gibbs specification is defined as the family of probability
kernels $(\gamma_{\Lambda}^{'})_{\Lambda\subset\subset V}$ from $(\Omega^{\nabla}, \mathcal{T}_{\Lambda}^{\nabla})$ to $(\Omega^{\nabla},\mathcal{F}^{\nabla})$ such that \begin{equation}\label{3.9}\int F(\rho)\gamma_{\Lambda}^{'}(d\rho | \zeta)=\int F(\nabla\varphi)\gamma_{\Lambda}(d\varphi | \omega) \end{equation} for all bounded $\mathcal{F}^{\nabla}$-measurable functions $F$, where $\omega\in \Omega$ is any height-configuration with $\nabla\omega =\zeta$.
\end{definition}

Using the sigma-algebra $\mathcal{T}_{\Lambda}^{\nabla}$, this is now a proper and consistent family of probability
kernels, i.e.
\begin{equation}\label{3.10} \gamma^{'}_{\Lambda}(A | \zeta)=1_{A}(\zeta)
\end{equation}
for every $A\in \mathcal{T}_{\Lambda}^{\nabla}$ and
$\gamma^{'}_{\Delta}\gamma^{'}_{\Lambda}=\gamma_{\Delta}^{'}$ for any finite volumes $\Lambda, \Delta\subset V$ with $\Lambda\subset \Delta$. The proof
is similar to the situation of regular (local) Gibbs specifications (\cite{Ge}, Proposition 2.5).

Let $\mathcal{C}_b(\Omega^{\nabla})$ be the set of bounded functions on $\Omega^{\nabla}$. Gradient Gibbs measures will
now be defined in the usual way by having their conditional probabilities outside finite regions prescribed by the gradient Gibbs specification:

\begin{definition}\label{3.2} A measure $\nu\in \mathcal{M}_1(\Omega^{\nabla})$ is called a gradient Gibbs measure (GGM) if it satisfies the DLR equation
\begin{equation}\label{3.11} \int \nu(d\zeta)F(\zeta)=\int\nu(d \zeta)\int \gamma_{\Lambda}^{'}(d\tilde{\zeta} | \zeta) F(\tilde{\zeta})
\end{equation}
for every finite $\Lambda\subset V$ and for all $F\in \mathcal{C}_{b}(\Omega^{\nabla})$. The set of gradient Gibbs measures will be denoted by $\mathcal{G}^{\nabla}(\gamma)$.
\end{definition}

\subsection{Translation-invariant gradient Gibbs measures}

In this subsection we construct gradient Gibbs measures for period 4 height-periodic boundary laws on the Cayley tree of order $k\geq 2$.

\begin{pro}\label{nup1}\cite{HK}  Probability distributions
$\mu^{(n)}(\sigma_n)$, $n=1,2,\ldots$, in (\ref{33})  are
compatible iff for any $x\in V\setminus\{x^0\}$ the following
equation holds:
\begin{equation}\label{nu5}
{\bf h}^*_x=\sum_{y\in S(x)}F({\bf h}^*_y,\theta).
\end{equation}
Here, $\theta=\exp(J\beta ),$ ${\bf
h}^*_x=(h_{i,x}-h_{0,x}+\ln\frac{\nu(i)}{\nu(0)},\, i\in \emph{Z}_0)$ and the  function $F(\cdot,\theta ): \, \emph{R}^{\infty}
\to  \emph{R}^{\infty}$ is $F({\bf h},\theta)=(F_{i}({\bf
h},\theta), \, i\in \emph{Z}_0)$, with
$$F_i({\bf h}, \theta )=\ln\frac{\nu(i)}{\nu(0)}
+\ln{\theta^{|i|}+\sum \limits_{j\in \emph{Z}_0}\theta^{|i-j|}\exp(h_j)\over 1+\sum \limits_{j\in \emph{Z}
_0}\theta^{|j|}\exp(h_j)},$$ ${\bf h}=(h_i, \, i\in \emph{Z}_0).$
\end{pro}

Assume ${\bf h}_x={\bf h}=(h_i,\, i\in \emph{Z}_0)$ for any $x\in V.$
In this case we obtain from  (\ref{nu5}):

\begin{equation}\label{11}
z_i=\frac{\nu(i)}{\nu(0)}\left({\theta^{|i|}+
\sum_{j\in  \emph{Z}_0}\theta^{|i-j|}z_j
\over
1+\sum_{j\in \emph{Z}_0}\theta^{|j|}z_j}\right)^k,
 \end{equation}
where $z_i=\exp(h_i), \ \ i\in \emph{Z}_0$.

Let $\mathbf z(\theta)=(z_i=z_i(\theta), i\in \emph{Z}_0)$ be a solution to (\ref{11}).   Denote
\begin{equation*}\label{lr}
l_i\equiv l_i(\theta)=\sum_{j=-\infty}^{-1}\theta^{|i-j|}z_j, \ \
r_i\equiv r_i(\theta)=\sum_{j=1}^{\infty}\theta^{|i-j|}z_j, \ \ i\in \emph{Z}_0.
\end{equation*}
It is clear that each $l_i$ and $r_i$ can be a finite positive number or $+\infty$. We shall consider all possible cases.

 Clearly, a solution $\mathbf z=(z_i, i\in \emph{Z}_0)$ to (\ref{11}) defines a tree-indexed Markov chain iff $r_0+l_0<+ \infty$ (see \cite{HK}).

Let $\nu(i)=1$ for any $i\in \emph{Z}$ then we consider the solutions of (\ref{11}) with $l_0<+\infty$ and $r_0<+\infty$.

Put $u_i=u_0\sqrt[k]{z_i}$ for some $u_0>0$. Then (\ref{11}) can be written as

\begin{equation}\label{45}
u_i=C\left( \sum_{j=1}^{+\infty}\theta^ju_{i-j}^k+u_i^k+ \sum_{j=1}^{+\infty}\theta^ju_{i+j}^k\right), \ \ i\in \emph{Z}.
\end{equation}

\begin{pro}\cite{HK}
A vector $\mathbf u=(u_i,i\in \emph{Z})$, with $u_0=1$,  is a solution to (\ref{45}) if and only if for $u_i \ \ (=\sqrt[k]{z_i})$ the following holds
\begin{equation}\label{V}
u_i^k={u_{i-1}+u_{i+1}-\tau u_i\over u_{-1}+u_{1}-\tau}, \ \ i\in \emph{Z},
\end{equation}
where $\tau=\theta^{-1}+\theta$.
\end{pro}

By this Lemma we have
\begin{equation}\label{1lr}
1+l_0+r_0={\theta-\theta^{-1}\over u_{-1}+u_1-\tau}.
\end{equation}

 Equations of system (\ref{11}) for $i=-1$ and $i=1$ are satisfied independently on values of $u_{-1}$ and $u_1$ and
the equation (\ref{V}) can be separated to the following independent recurrent equations
\begin{equation}\label{L}
u_{-i-1}=(u_{-1}+u_1-\tau)u_{-i}^k+\tau u_{-i}-u_{-i+1}, \end{equation}
\begin{equation}\label{99}
u_{i+1}=(u_{-1}+u_1-\tau)u_{i}^k+\tau u_{i}-
u_{i-1}, \end{equation}
where $i\geq 1$, $u_0=1$ and $u_{-1}$, $u_{1}$ are some initial numbers (see again \cite{HK}).

So, if $u_i$ is a solution to (\ref{99}) then $u_{-i}$ will be a solution for (\ref{L}). Hence we can consider only
 equation (\ref{99}).

Let's consider the periodic solutions of (\ref{V}) i.e., we describe solutions of (\ref{V}) which have the form
\begin{equation}\label{up}
u_n=\left\{ \begin{array}{lll}
1, \ \ \mbox{if} \ \ n=2m,\\[2mm]
a, \ \ \mbox{if} \ \ n=4m-1, \ \ m\in \emph{Z}\\[2mm]
b, \ \ \mbox{if} \ \ n=4m+1,
\end{array}
\right.
\end{equation}
where $a$ and $b$ some positive numbers.
In this case (\ref{99}) is equivalent to the following system of equations
\begin{equation}
\label{ab}
\begin{array}{ll}
(a+b-\tau)b^k+\tau b-2=0\\[2mm]
(a+b-\tau)a^k+\tau a-2=0.
\end{array}
\end{equation}

We describe positive solutions of (\ref{ab})

\textbf{Case $a\neq b$}. We multiply the first equation of (\ref{ab}) by $a^k$ and the second equation of (\ref{ab}) by $b^k$.
And after that, subtract the first equation from the second and we obtain the following equation:
\begin{equation*} \tau ab(a^{k-1}-b^{k-1})-2(a^k-b^k)=0\end{equation*}
 Dividing both sides by $a-b$ we get
 \begin{equation*} (a^{k-1}+a^{k-2}b+...+a^2b^{k-3}+ab^{k-2})(\tau b-2)-2b^{k-1}=0.
\end{equation*}
Put $x:=\frac{a}{b}$, then the last equation can be written as
 \begin{equation*} (x^{k-1}+x^{k-2}+...+x^2+x)(\tau b-2)-2=0.
\end{equation*}

If $\tau b-2\leq 0$ then $(x^{k-1}+x^{k-2}+...+x^2+x)(\tau b-2)-2<0,$ i.e., there is not any solution $(a,b)$
of (\ref{ab}) such that $a\neq b.$

Let $\tau b-2> 0$ then for any positive fixed $b$ we consider the following polynomial $P_b(x):=(x^{k-1}+x^{k-2}+...+x^2+x)(\tau b-2)-2.$ For $x>0$ it's easy to check that $P^{'}_b(x)>0$ and $P_b(0)<0$, $\lim\limits_{x\rightarrow\infty}P_b(x)>0$.
Thus, $P_b(x)$ has exactly one positive solution. If $\tau\beta=2+\frac{2}{k-1}$ then there is not any solution $(a,b)$
to (\ref{ab}) such that $a\neq b.$ In other cases, from $P_b(1)\neq 0$ for any positive $b$ there exists a unique $a(b)\neq b$ such that $(a,b)$ is solution
to (\ref{ab}).

\textbf{Case $a=b$}. In this case it is sufficient to consider one of equations of $(\ref{ab})$, i.e.,
$$2a^{k+1}-\tau a^{k}+\tau a-2=0.$$
Last equation has the solution $a=1$ independently on the parameters $(\tau, k).$ Dividing both sides by $a-1$ we get
\begin{equation}\label{4.18} Q(a):=2a^k+(2-\tau)(a^{k-1}+a^{k-2}+...+a)+2=0\end{equation}
By definition of $\tau$ we get $\tau\geq 2$ i.e., From $2-\tau< 0$ and Descartes' rule of signs,
$Q(a)$ has at most two positive roots. Since $Q^{'}(a)=2ka^{k-1}+(2-\tau)((k-1)a^{k-2}+...+2a+1)$ and $Q^{'}(0)<0$, $Q^{'}(\infty)>0$
 there is a unique $a_{c}$ such that $Q^{'}(a_{c})=0$. Consequently, if $\tau_c:=Q(a_{c})<0$ then the polynomial $Q(a)$ has exactly two positive solutions. Let $Q(a_{c})=0$ then $Q(a)$ has exactly one positive solution. Finally, if $Q(a_{c})>0$ then the polynomial $Q(a)$ has not any positive solution.

\begin{thm}\label{new}(Theorem 4.1, Remark 4.2 in \cite{cp}). Let $l$ be any spatially homogenous period-$q$ height-periodic boundary law to a tree-automorphism invariant gradient interaction potential on the Cayley tree. Let $\Lambda\subset V$ be any finite connected set and let $\omega\in \Lambda$ be any vertex. Then the measure $\nu$ with marginals given by
\begin{equation}\label{88} \nu(\eta_{\Lambda\cup\partial\Lambda}=\zeta_{\Lambda\cup\partial\Lambda})=Z_{\Lambda}
\left(\sum_{s\in \emph{Z}_q}\prod_{y\in\partial\Lambda}l\left(s+\sum_{b\in\Gamma(\omega, y)}\zeta_b\right)\right)\prod_{b\cap\Lambda\neq\emptyset}Q(\zeta_b),\end{equation}
\end{thm}
From above results and Theorem \ref{new} we can conclude the following theorems:
\begin{thm}\label{rozikov1} Let $k\geq 2$ and $a=b$. For the SOS-model (\ref{nu1}) on the $k$-regular tree, with parameter $\tau=2\cosh(\beta)$ there
numbers $\tau_c>0$ such that the following assertions hold:
\begin{enumerate}
  \item If $\tau<\tau_c$ then there is a unique GGM corresponding to nontrivial period-3 height-periodic boundary laws of the type (\ref{up}) via Theorem (\ref{88})
  \item At $\tau=\tau_c$ there are exactly two GGMs corresponding to a nontrivial period-3 heightperiodic boundary law of the type (\ref{up}) via Theorem
  \item For $\tau>\tau_{c}$ there are exactly three such (resp. one) Gradient GMs.
\end{enumerate}
\end{thm}
\begin{thm}\label{rozikov} Let $k\geq 2$ and $a\neq b$. For the SOS-model (\ref{nu1}) on the $k$-regular tree, with parameter $\tau=2\cosh(\beta)$ the following assertions hold:
\begin{enumerate}
  \item For any positive fixed $b$, if $\tau\leq \frac{2}{b}$  then there is no any Gradient Gibbs Measure (GGM) corresponding to nontrivial period-3 height-periodic boundary laws of the type (\ref{up}) via Theorem (\ref{88}).
  \item For any positive fixed $b$, if $\tau>\frac{2}{b}$  then there is a unique GGM corresponding to nontrivial period-3 height-periodic boundary laws of the type (\ref{up}) via Theorem (\ref{88}).
\end{enumerate}
\end{thm}

\section*{ \textbf{Acknowledgements}}
We are deeply grateful to Professor U.A.Rozikov for the attention to our work and useful suggestions.
We thank the referee for many helpful comments.

\end{document}